\documentstyle[12pt]{article}
\newlength{\halfpage}
\begin{document}

\noindent {\bf Note on a Comment by Edward L. Wright}

\noindent {F. Hoyle\footnote{102 Admirals Walk, West Cliff Road,
Bournemouth BH2 5HF, UK.}, G. Burbidge\footnote{UCSD, Department of
Physics, Center for Astrophysics and Space Sciences, La Jolla, CA
92093-0111, USA.}, and J.V. Narlikar\footnote{IUCAA, Post Bag 4, Pune
411 007, INDIA.}

\rm
In a recent paper Edward L. Wright (1994) has criticized both a paper of ours
and the {\em Monthly Notices}
which published it (Hoyle, et al. 1994) on the
original grounds that neither we nor the journal took adequate notice of
his views!  He is concerned with Sections 3 and 4 of our second paper in
the series in which we have developed the quasi-steady state
cosmological model and have shown that it is a viable alternative to the
popular big bang model.  We believe that his criticisms of our Section 3
have already been discussed adequately in our paper and we have nothing
further to add.  His discussion of our Section 4, while not seeming to
differ significantly from our calculations technically, differs in
relation to a radiosource survey by Allington-Smith (1982).

In that survey, Allington-Smith attempted identifications of 59 sources
with flux values between 1 and 2 {\em Jy} taken over a limited strip of
the sky from the Bologna survey at 408 MHz.  We were concerned in our
paper with radiogalaxies, and of these 59 sources about 32 would be
expected to be radiogalaxies.  Our paper predicted about 8 should show
identifications at magnitudes of $\sim +20$ to $+21$, and the remaining
24 would show identifications at about magnitude $+25$.  Since
Allington-Smith's attempted identifications went down to $r$ magnitude
about $+23$, our expectation was that the 24 cases would be found as
so-called empty fields.  But things turned out the opposite way round,
about 8 empty fields and 24 rather clear-cut identifications at
magnitudes typically around $+20$.

We had been aware of this survey, at the time of our paper, but had
discounted it, partly because of the small number of sources involved,
about 32 radiogalaxies out of 5000 or more with $F \geq 1 Jy$ over the
whole sky, and partly because there seemed to be an incompleteness in
its redshift content (McCarthy 1993) Wright's view however, is that the
survey should be considered to give a complete sample of all 408 MHz
sources with $1 \leq F 2 \leq Jy$.  If this view is accepted, then the
question of whether our theory can match the observed source counts as a
function of $F$ requires further consideration.  We show in what follows
that the issue is a tactical one, in no way involving the basic
properties of our cosmological model.

The form in which we consider the quasi steady-state model had a scale
function
\begin{eqnarray}
S(t) & = & \exp \frac{t}{P} \left( 1 + 0.75 \cos \frac{2 \pi t}{Q}
\right) \\ \nonumber
\end{eqnarray}
\noindent in the metric.
\begin{eqnarray}
ds^{2} & = & dt^{2} - \frac{S^{2}(t)}{S^{2}(t_{\circ})} \left[ dr^{2} +
r^{2} ( d \theta^{2} + \sin^{2} \theta d \phi^{2} ) \right] , \\
\nonumber
\end{eqnarray}
\noindent where $t_{\circ}$ is the present moment of time.  The
characteristic period of oscillation of the second factor of (1) was
taken as $Q = 40 \times 10^{9}$ years, while $P \gg Q$.  The model has a
slow expansion due to the factor $\exp \; t/P$ superposed on the
oscillations.  Numerical values were given in former papers subject to
the choice $T_{\circ} = 0.85 Q$, or $t_{\circ} = 0.85$ when Q is used as
the unit of time.

We took the occurrence of radio sources to be always proportional to the
cosmological density, while here we modify this extremely simple
postulate by requiring the occurrence of powerful sources to be
concentrated in the half of each cycle centered around the minima of
$S(t)$.  This accords with the view we have expressed on several
occasions more recently than our Monthly Notices paper cited above,
namely that violent activity in our model tends to be centered around
the minima of the oscillatory cycles.  On this view we expect sources of
very high radio luminosity to be more concentrated towards the minima
than sources of comparatively low luminosity.  Explicitly, we now
consider a rather simple situation in which there are just three
luminosities:
\begin{description}
\item[I] sources with $L = 10^{26}WHz^{-1}$,
\item[II] sources with $L = 10^{27} WHz^{-1}$,
\item[III] sources with $L = 10^{29}WHz^{-1}$.
\end{description}

\noindent A fourth class with $L = 10^{28}WHz^{-1}$ could be included
and would give extra scope to the model.  But since such a fourth class
is not essential we omit it for simplicity.

Sources of Type I we consider to occur uniformly at all times, as in our
former discussion.  Sources of Type II are to occur uniformly as before
for $0.22 \leq t \leq 0.78$ but not for $t \geq 0.78$, and sources of
Type III occur uniformly as before for $0.30 \leq t \leq 0.70$ but not
for $t \geq 0.70$.  And the relative occurrence rates at times when all
three types occur as given by I:II:III = 5000:1000:1.  Very powerful
sources are thus infrequent.  The resulting integral source counts as a
function of flux $F$ are given by the points of Figure 1, which may be
compared with the observed counts shown in Figure 2 (Fig 2 of Hoyle et
al. 1994a).  Only sources from
the present half-cycle $0.5 \leq t \leq 0.8$ contribute effectively to
Figure 1 and all have redshifts.  Integral source counts have been used
in Figure 1, otherwise the approximation of sources of Type II and III
being assumed to begin abruptly at $t \leq 0.78$ and $t \leq 0.70$
respectively (rather than as continuous transitions) would lead to
artificial distortions.

Only types I and II contribute down to the level of the 3CR catalog at $F =
10 Jy$, in a ratio of about 3:1.  The Type I sources have low redshifts
of $\sim 0.03$, while the Type III sources have redshifts of $\sim 1.4$.
 At $F \simeq 1 \; Jy$, however, the count is dominated by Type II
sources with redshifts of $\sim 0.5$.  These are the cause of the
notorious rise of the counts as $F$ decreases from $10 \; Jy$ to $\sim 1
\; Jy$.

If it be said that the above model is analogous to the way in which
Big-Bang cosmology seeks to grapple with the source count problem, our
reply is that our choice  of parameters can be understood in relation to
the phases of the oscillatory cycles in $S(t)$, whereas there is no
similar situation in Big-Bang cosmology, where radio sources are
required to change their behavior quite drastically just about now.
Subsequent to the synthesis of the light elements in Big-Bang cosmology,
the density of the Universe is supposed to have fallen by some thirty
powers of ten.  And yet radio source evolution is required to have
occurred in the last decline of the density by a factor only $\sim 2$.
Why?

In his concluding remarks Wright appears to have been completely carried
away (as he was in his original referee's report).  He states that ``the
data disprove the QSSC''.

As we have said elsewhere the QSSC is based on a fundamentally new idea
involving matter creation in an oscillatory mode imposed on an expanding
universe (Hoyle et al. 1993, 1994a,b).  The exact values of the free
parameters of the theory are not known.  The points in dispute are
related to the numerical values we have chosen when we compare with
observational details.  Suitable choices can and do make the theory
compatible with the observations, i.e. it is obvious that the theory is
not disproved by Wright's analysis of the data.

\noindent {\bf REFERENCES}

\begin{description}
\item[] Allington-Smith, J.R. 1982, MNRAS, 199, 611
\item[] Hoyle, F., Burbidge, G. and Narlikar, J.V. 1993, ApJ, 410, 437
\item[] Hoyle, F., Burbidge, G. and Narlikar, J.V. 1994a, MNRAS, 267,
1007
\item[] Hoyle, F., Burbidge, G. and Narlikar, J.V. 1994b, A\&A, 289, 729
\item[] McCarthy, P.J. 1993, ARA\&A, 31, 639
\item[] Wright, E.L. 1994, preprint.
\end{description}

\pagebreak

\noindent {\bf Figure Captions}

\begin{description}
\item[Fig 1] Theoretical curve based on the model described here for the
counts of radio sources
\item[Fig 2] Observed counts of radio sources at 408 MHz and 1.4 GHz,
with the latter restricted to flux values greater than $\sim 10^{-2}
Jy$, redrawn from Kellermann \& Wall (1987).  The counts are normalized
to a uniformly filled static Euclidean universe.
\end{description}
\end{document}